\documentclass{llncs}
\usepackage[utf8]{inputenc}

\usepackage{graphicx}
\usepackage{subcaption}
\usepackage{siunitx}
\DeclareSIUnit{\sample}{S}
\usepackage{mathtools}

\usepackage{amssymb}

\newcommand\nSub{7 }

\title{Synthesizing dynamic MRI using long-term recurrent convolutional networks}
\author{Frank Preiswerk\inst{1} \and Cheng-Chieh Cheng\inst{1} \and Jie Luo\inst{1,2} \and Bruno Madore\inst{1}}
\institute{Brigham and Women's Hospital, Harvard Medical School, USA\\
\and Graduate School of Frontier Sciences, The University of Tokyo, Japan}

\date{February 2018}

\begin{document}

\maketitle

\begin{abstract}
A method is proposed for converting raw ultrasound signals of respiratory organ motion into high frame rate dynamic MRI using a long-term recurrent convolutional neural network. Ultrasound signals were acquired using a single-element transducer, referred to here as `organ-configuration motion' (OCM) sensor, while sagittal MR images were simultaneously acquired. Both streams of data were used for training a cascade of convolutional layers, to extract relevant features from raw ultrasound, followed by a recurrent neural network, to learn its temporal dynamics. The network was trained with MR images on the output, and was employed to predict MR images at a temporal resolution of 100 frames per second, based on ultrasound input alone, without  any further MR scanner input. The method was validated on \nSub subjects.
\end{abstract}

\section{Introduction}
Ultrasound (US) and Magnetic Resonance Imaging (MRI) signals are highly complementary. MRI is based on magnetic and RF fields and can achieve diversified soft-tissue contrasts, while US imaging is based on longitudinal pressure waves and offers a high temporal resolution, convenient and relatively low cost approach to diagnostic imaging. Efforts have been made to combine these two very different modalities, for US-MRI image fusion \cite{Petrusca_2013}, as well as prospective motion compensation in MRI \cite{Feinberg_2010}, using brightness mode (B-mode) ultrasound. A potentially useful idea in the context of image-guided intervention would be to learn the appearance of free-breathing MRI images during a training stage, then estimate them later on when MRI scanning may not be available anymore, for example after the patient left the MRI suite. Whether on the same day or a different day, the ability to generate MRI contrast based solely on US signals alone would be helpful as the patient proceeds to other diagnostic and/or therapy device(s), to continue generating MRI-like images even as the patient lies in a positron-emission tomography (PET) scanner or a radiotherapy device, for example. To this end, 
the approach introduced in \cite{Preiswerk_2016} and the publicly-available software\footnote{https://github.com/fpreiswerk/OCMDemo} was considerably expanded here to allow the rapid synthesizing of MRI contrast using a long-term recurrent convolutional network inspired from the video-recognition work in \cite{Donahue_2014}.

An MR-compatible single-element ultrasound transducer \cite{Schwartz_2013} and a 3D-printed capsule, collectively referred to here as an ’organ-configuration motion’ (OCM) sensor, acquired amplitude mode (A-mode) US signals of respiratory organ motion. In contrast to the conventional 2D spatial interpretation of US signals through delay-and-sum beamforming, the OCM's A-mode signals were not spatially encoded but provided a high temporal resolution signature of abdominal configuration, sensitive over a region in the area of sensor placement. Fast OCM signals (100 fps) can be correlated with slower-rate MRI acquisitions (1 fps), to estimate fast synthetic MR images of respiratory organ motion at the rate of the OCM signals (100 fps).
This could be done using kernel density estimation (KDE) \cite{Nadaraya_1964,Watson_1964} to model this relationship in a non-parametric way, as data is acquired during online learning, as proposed in  \cite{Preiswerk_2015,Preiswerk_2016}. KDE is well suited for online learning, because there is no separation into training and inference stage. However, this comes at computational cost, as the time complexity at inference depends on the size of the dataset. In \cite{Preiswerk_2016}, an image reconstruction time of 45 ms for a single 2D MR image was reported using such KDE approach, on a relatively-small database accumulated over \SI{2}{\minute} of hybrid OCM-MRI data. Furthermore, the inter-fraction variability of OCM signals was reported to be significant, which would presumably prevent any removal/re-attachment of an OCM probe, and confuse the KDE-based processing. As a result, any scenario involving the use of MRI+OCM data acquired on a given day to supplement, for example, radiotherapy treatments performed on a different day could not be considered, as the removal and re-attachment of the sensor days later would destroy the ability to generate accurate MR images from OCM signals. Lastly, due to the curse of dimensionality being a limiting factor in kernel methods, a small subset of depth values had to be pre-selected in the OCM traces in \cite{Preiswerk_2016}, as a trade-off between information vs. dimensionality of the data.  
Recently, artificial neural networks have become state-of-the-art models for computer vision (CV) and natural language processing (NLP) \cite{LeCun_Bengio_Hinton_2015}. Feed-forward architectures, most notably convolutional neural networks (CNNs) \cite{Lecun_1998} are used to automatically extract hierarchical features from (labeled) data, while recurrent networks (RNNs), typically based on long-short term memory (LSTM) units \cite{Hochreiter_1997}, allow temporal structures to be learned from data.
We propose to use a combined CNN-LSTM model, called a long-term recurrent convolutional network (LRCN) \cite{Donahue_2014}, to learn the relationship between OCM sensor data and fully reconstructed MR images end-to-end. Our method improves on all the aforementioned challenges associated with KDE; By directly learning a mapping between OCM signals and MR images, the computational cost of image reconstruction is shifted from inference time to the training stage. Hence, the computational cost of image reconstruction becomes independent of the training set size. Our approach can therefore, in principle, be scaled to estimating several planes at once, i.e., 4D-MRI, at a high temporal rate. Our pre-processing step, closely related to Doppler processing, makes OCM signals more robust against signal changes that have little to do with physiological motion, and more 
to do with inconsequential details on exact sensor placement and/or anatomy. As a consequence, the Doppler-like pre-processing may help avoid registration steps when removing and re-attaching OCM sensors. Lastly, the curse of dimensionality is defeated since, unlike kernel methods, the proposed method does not rely on a high-dimensional similarity measure to be evaluated between any new OCM signal and all signals from the training set.

\section{Materials and Methods}

\begin{figure}[t]
    \centering
    \begin{subfigure}[b]{0.42\textwidth}
        \includegraphics[width=\textwidth]{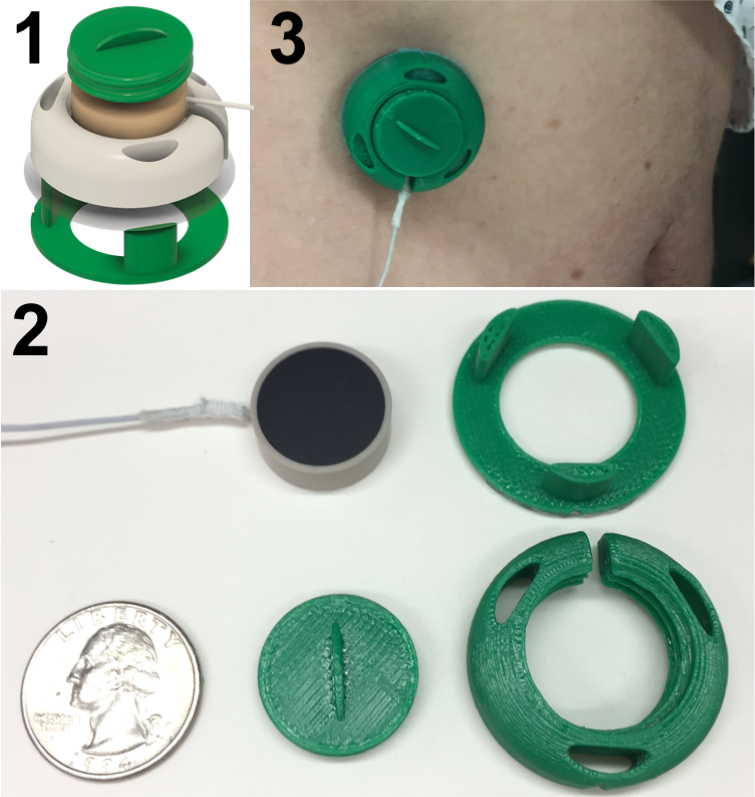}
        \caption{}
        \label{fig:setup_transducer}
    \end{subfigure}
   \qquad
    \begin{subfigure}[b]{0.34\textwidth}
        \includegraphics[width=\textwidth]{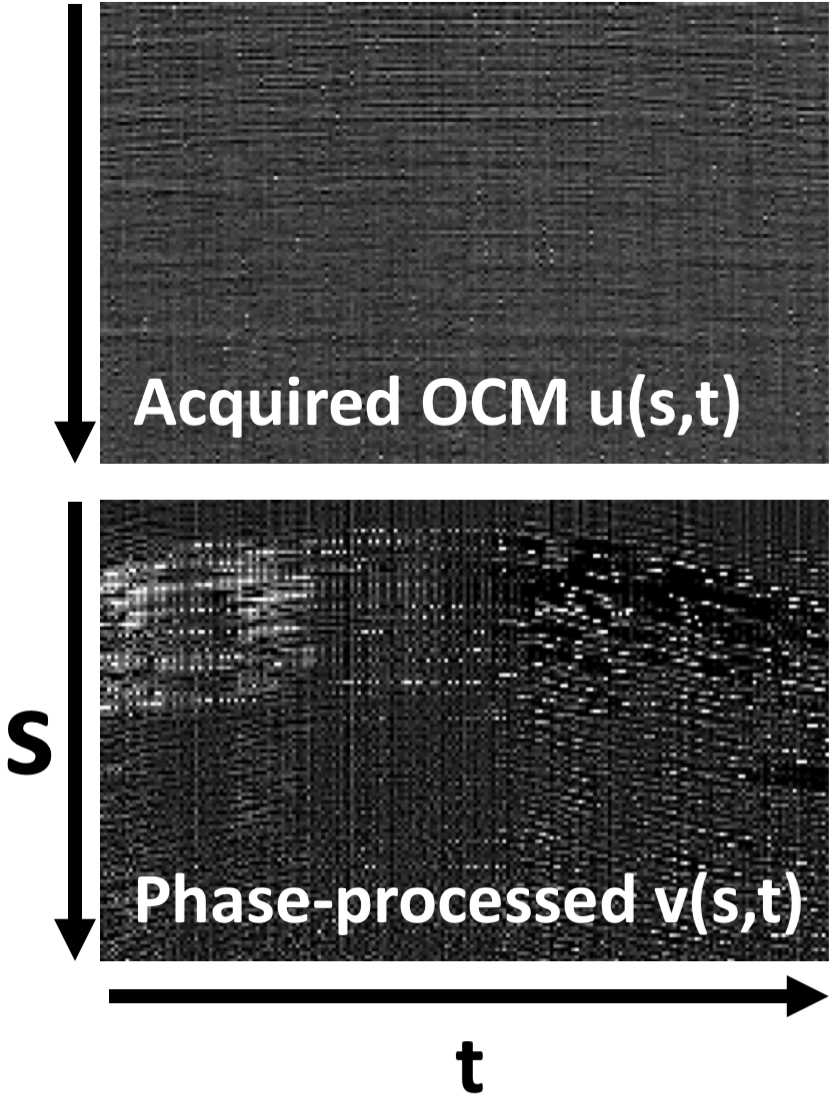}
        \caption{}
        \label{fig:OCM_data}
    \end{subfigure}
    \caption{a) 3d model rendering (1) and Individual parts (2) of an OCM sensor. The US transducer (a 1 Mhz version is depicted in (a)) was fitted into a 3d-printed capsule of our own design (green parts), which housed water-based gel for acoustic coupling and allowed for the pressure onto the skin to be adjusted by twisting the screw-like lid (3). Two-sided tape on the bottom was used for adhesion on the skin. b) Visualization of unprocessed signals $u$ and phase-processed versions $v$ over a 3 s window. Respiratory motion is more pronounced in phase-processed signals (bright/dark pixels correspond to traces acquired during inspiration/expiration, respectively).}\label{fig:hardware_setup}
\end{figure}

Hybrid OCM-MRI data were acquired on \nSub subjects following informed consent using an IRB-approved protocol. Scanning was performed on a Siemens Verio 3T system, using a T1-weighted spoiled gradient echo MRI sequence with two-fold parallel imaging acceleration and 5/8 partial-Fourier acceleration. The US transducer at the heart of the OCM sensor was either a 5MHz (subjects 1-4) or 1MHz (subjects 5 and 6) MR-compatible transducer (Imasonics). The transducer was enclosed in a custom 3d-printed capsule that allowed for quick and easy attachment to the skin, regulation of pressure through a screwable lid (see Figure \ref{fig:hardware_setup}), and retention of water-based US gel for acoustic coupling. The 1MHz transducer employed in later subjects achieved greater signal penetration; nevertheless, both 5MHz and 1MHz OCM signals appeared equally appropriate for our purpose.

The OCM data acquisition was synchronized with the scanner's repetition rate, TR = \SI{10}{\milli\second}, using dedicated hardware and minor modifications to the MRI pulse sequence: At the beginning of each TR interval, the scanner was programmed to generate an optical synchronization pulse, which was then converted to a TTL voltage pulse using dedicated hardware. These pulses were used to trigger the OCM acquisition, at the rate of exactly one OCM trace acquisition per TR interval, thus precisely synchronizing the MRI and OCM streams of data. The purpose of such synchronization was two-fold: to allow MRI and OCM data to be unambiguously located on a common time axis, and to avoid the OCM sensor being fired during an MRI acquisition window, which would have caused artifacts in the MRI images. A total of 60 k-space lines and corresponding OCM signals were acquired per image. Individual OCM traces $u$ were sampled at $f_s = \SI{100}{\mega\sample\per\second}$ for $t_s = \SI{200}{\micro\second}$, yielding $D = f_s \cdot t_s = 20 e^3$ samples per trace. The window from index 1000 to 8000 was retained for further processing, and downscaled to $d=560$ samples. MR images of the breathing liver in the sagittal plane were acquired at a rate of \SI{0.85}{fps}. Figure \ref{fig:hardware_setup} gives an overview of the OCM sensor and data.

\subsubsection{Preprocessing of OCM signals and MR images:} 
Raw (magnitude) OCM signals $u(s,t)$ are highly sensitive to physiological motion along $t$ (the repeat index, as OCM traces are repeatedly acquired every TR = \SI{10}{\milli\second}), but unfortunately, they tend to also prove highly sensitive to mostly unimportant details along $s$ (the sampling index) relating to sensor placement and underlying anatomy. In the process to separate the former from the latter, OCM signals were first transformed into a complex entity:
\begin{equation}
    \hat{u}(s,t) = \mathcal{F}_s^{-1}(\Omega(\mathcal{F}_s(u(s,t)))) = |u(s,t)| (\cos \theta(s,t) + i \sin \theta(s,t)),
\end{equation}
where $\mathcal{F}_s$ is the discrete Fourier transform along $s$, and $\Omega$ is a Fermi filter that cancels negative as well as very high frequencies ($>10 \cdot f_0$, where $f_0$ is the transducer center frequency). In analogy with Doppler ultrasound, we shall now consider $\theta(s,t)$, the complex angle of $\hat{u}(s,t)$ for further analysis. Variations along $s$ have more to do with the object itself rather than how the object moves; for this reason the signal evolution along $t$, i.e., from trace to trace, was more closely linked to internal organ motion than variations along $t$. In particular, from $\theta(s,t)$, speed can be computed according to
\begin{eqnarray}
    v(s,t) &=& \alpha \cdot \frac{d\theta(s,t)}{dt} = \alpha \cdot \frac{\theta(s,t) - \theta(s,t-1)}{2},
\end{eqnarray}
with $\alpha = \frac{0.5 \cdot \lambda}{360}$, where $\lambda$ is the wavelength in mm. Figure \ref{fig:OCM_data} visualizes $u$ and $v$. We denote the vector of signals of a single timestep $t$, over the whole signal depth $s = \{1, \ldots, d\}$, as $\mathbf{v}(t) \vcentcolon= [v(1,t), \ldots, v(d,t)]^T$. 
For further processing, OCM signals were rearranged as $X_t \vcentcolon= [\mathbf{v}(t{-}n{+}1), \ldots, \mathbf{v}(t)]$, combining the most recent signal history of length $n=300$ (\SI{3}{\second}) in the form of a 2d image patch. This format proved well suited as input to the neural network described in the next section.
Instead of explicitly modeling all pixels of the MR image domain (i.e., the model output dimension), we exploit correlations between pixels by compressing the images first, using Principal Component Analysis (PCA); 10 principal components are retained and used as target variables $\mathbf{y}_t$ for the neural network. This compression from size 192\,px $\times$ 192\,px into a vector of 10 principal components for each image allows to significantly reduce the number of parameters in our model, at the cost of an acceptable loss of high-frequency image content.

\subsection{Network architecture}

In \cite{Preiswerk_2016}, KDE is used to compute the expectation of unknown MR images $I_t$, given new OCM signals $X_t$ and a database of previously seen data $D_t = \{I_\tau, U_\tau | \tau < t\}$,
\begin{equation}
    \mathbb{E}_{I \sim p(I | X)} [ I_t | X_t, D_t ]
    \label{eq:expectation}
\end{equation}
From a learning theory perspective, our motivation to replace KDE with a neural network to solve Equation \ref{eq:expectation} is guided by the following result from calculus of variations. We can view a neural network as any function $f$, granted the network is sufficiently powerful. Learning then becomes equivalent to choosing the best function according to the variational problem
\begin{equation}
    f^* = \underset{f}{\operatorname{argmin}} \,\, \mathbb{E}_{I, X \sim p_{data}} \, ||I_t - f(X)||^2,
    \label{eq:argmin}
\end{equation}
which has a solution at 
\begin{equation}
    f^*(X) = \mathbb{E}_{I \sim p_{data}(I|X)} [I].
\end{equation}
In the hypothetical case where infinitely many samples are available, Equation \ref{eq:argmin} implies that the mean squared error loss leads to an optimal estimate of Equation \ref{eq:expectation}, so long as $f^*$ is part of the class of functions we optimize over. In practice, of course, a limited amount of data is available, and regularization techniques are typically applied.
The major difference to non-parametric approaches, including KDE, is that a set of fixed model parameters is obtained. If the number of neurons is treated as a constant, the time complexity of a single prediction equals O(1), while a single prediction using KDE has complexity $\mathcal{O}(Nd)$, where $N$ is the number of OCM training samples in $D$, and $d$ is their dimensionality.

\begin{figure}[t]
    \centering
    \begin{subfigure}[b]{0.45\textwidth}
        \includegraphics[width=\textwidth]{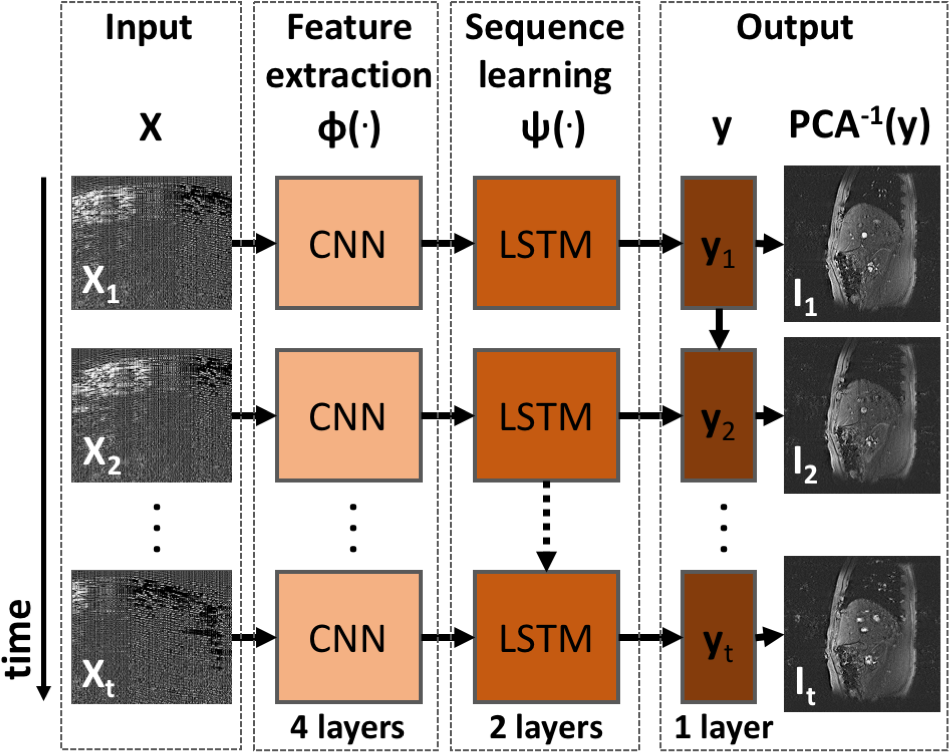}
        \caption{}
        \label{fig:LRCN_overview_overall}
    \end{subfigure}
    \quad
    \begin{subfigure}[b]{0.49\textwidth}
        \includegraphics[width=\textwidth]{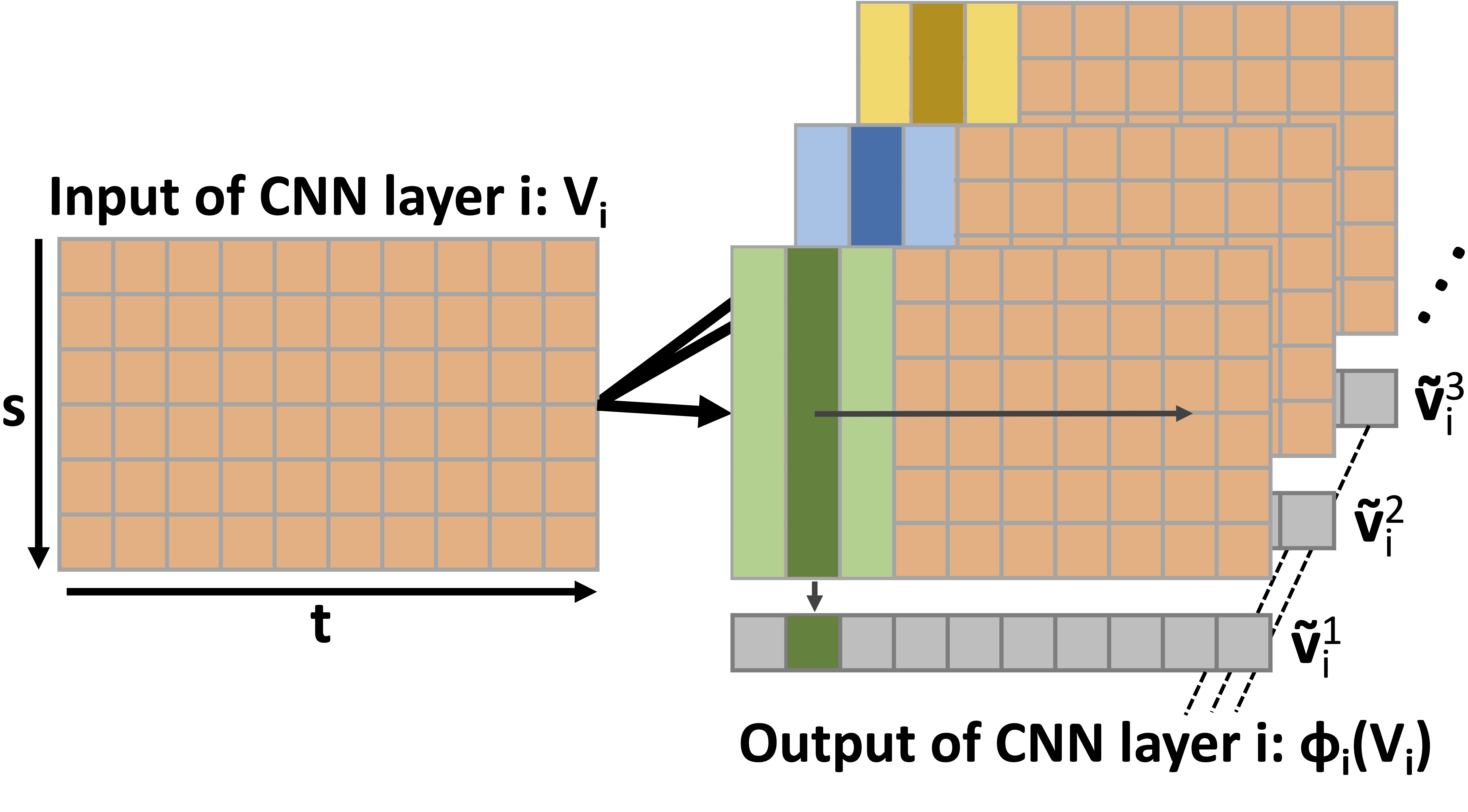}
        \vspace{0.23cm}
        \caption{}
        \label{fig:LRCN_overview_cnn}
    \end{subfigure}
    \caption{a) Overview of unrolled long-term recurrent convolutional network (LRCN) structure. 1-d convolutional layers extract image features from the input, and recurrent layers  learn the temporal evolution of transformed features, using LSTM units. A densely-connected layer maps to 10 PCA coefficient outputs, $\mathbf{y}_i$ used to reconstruct the final image $\hat{I}_t$. b) Detailed view of a 1-d convolutional layer. Convolutions are applied along the spatial dimension $s$ only, transforming all columns to a vector $\mathbf{v}$. Colors green, blue and yellow represent different feature maps, i.e. learned kernels. A dense output layer transforms the LSTM outputs into a vector of principal components, and the inverse PCA transform (PCA$^{-1}$) restores the final image.}\label{fig:LRCN_overview}
\end{figure}

Inspired by recent work in image captioning and related tasks in video analysis, a long-term recurrent convolutional network (LRCN) \cite{Donahue_2014} architecture is used to learn the mapping $f(\cdot)$ from signals $X_t$ to MR images, $f(X_t) = \mathbf{y}_t$. The network consists of convolutional layers, $\phi_\tau(\cdot)$, followed by recurrent layers $\psi_\upsilon(\cdot)$, $f_{\tau,\upsilon}(X_t) = \psi_\upsilon({\phi_\tau(X_t))}$, both with their respective set of parameters $(\tau, \upsilon)$. For brevity, we omit these parameters from here on. Figure \ref{fig:LRCN_overview_overall} shows an overall picture of the network.
The purpose of the convolutional layers is to extract features over the spatial dimension, $s$ (i.e., columns), from input signals $X_t$. To this end, 1-d convolutions are applied along $s$; each output feature map corresponds to one 1-d filter applied along $s$ to all columns of the input. Thus, the output of convolutional layer $\phi_i(\cdot)$ is a set of $k_i$ row vectors $V_i = [\{\mathbf{\tilde{v}}_i^1\}^T, \ldots,\{\mathbf{\tilde{v}}_i^{k_i}\}^T]$ (another image), each row being a convolved version of all columns in $V_{i-1}$ (see Fig. \ref{fig:LRCN_overview_cnn}). Down at the last convolutional layer $l$, $k_l=1$, so its output $V_i = \{\mathbf{\tilde{v}}_i\}^T$ represents one-dimensional encoding of the signal evolution over the $n$ time steps contained in $X_t$. It is now the task of the following recurrent layers to learn how this encoding evolves over time. Recurrent layer $i$ transforms its input according to $\psi_i(V_i,h_{t-1})$, where $h_{t-1}$ is its internal state from the previous time step. Through this recurrence, coupled with an internal memory state, recurrent units are able to learn from the arbitrarily distant past, if necessary. Long-short term memory (LSTM) units \cite{Hochreiter_1997} are used here in all recurrent layers. Finally, a densely-connected output layer at depth $L$ maps $V_{L-1}$ to final outputs $\mathbf{y} = g(W V_{L-1})$, with weight matrix $W$ and linear activation $g(\cdot)$. For all experiments, the network structure was set to 4 convolutional layers with 64, 32, 16 and 1 output channels, respectively, followed by 2 recurrent layers with 10 output channels each. Both convolutional and recurrent units use $tanh$ activations. The network architecture is depicted in Figure \ref{fig:LRCN_overview_cnn}. Not shown in the figure are average pooling operations (pool size 2) between all convolutional layers, as well as are dropout layers (rate 0.2) active during training on all convolutional layers. 

\section{Results and Discussion}
\begin{figure}[t]
    \centering
    \begin{subfigure}[b]{0.56\textwidth}
        \includegraphics[width=\textwidth]{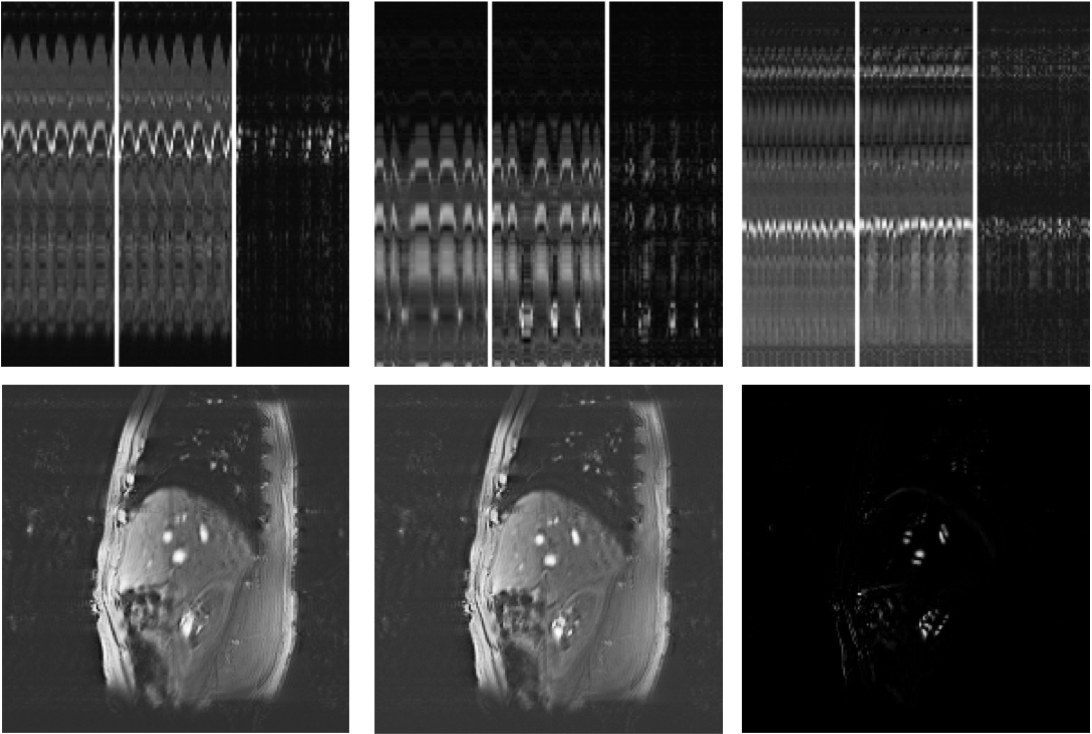}
        \caption{}
        \label{fig:MRI_comparison}
    \end{subfigure}
      \quad
    \begin{subfigure}[b]{0.40\textwidth}
        \includegraphics[width=\textwidth]{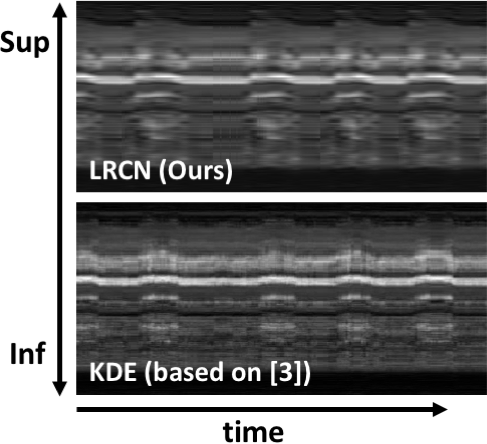}
        \caption{}
        \label{fig:KDE_comparison}
    \end{subfigure}
    \caption{a) Top row: M-mode display of all 50 test images of subjects 1-3. Reconstruction, ground-truth (in PCA space) and difference images side-by-side. Bottom row: Random image from subject 1 from test set, ground-truth (in PCA space) and difference image. b) Comparison of LRCN vs. KDE approach\cite{Preiswerk_2016}, where an average error of \SI{1}{pixel} was reported through manual validation by a radiologist. LRCN-based reconstructions are smoother but comparable.\label{fig:KDE_comparison_all}}
\end{figure}

Each of the \nSub datasets was separated into a training set of $\SI{60}{\second}$ (100 MR images of size 192 $\times$ 192\,px with 100 corresponding OCM signal histories of 300 $\times$ 560\,px) and a test set of $\SI{30}{\second}$ (50 MR images with OCM signals). For each subject, a separate LRCN model was trained on the training set and evaluated on the test set. The mean squared error loss function was employed to optimize the $28,471$ trainable parameters of each of the 7 networks, using the Adam optimizer (learning rate 0.001, $\beta_1$=0.9, $\beta_2$=0.999) over 1000 epochs. Training time was below \SI{5}{\minute} per dataset on an NVIDIA Titan X GPU. Code and sample data is available online\footnotemark[1]. Figure \ref{fig:MRI_comparison} compares MRI reconstructions from the test set with their ground-truth. High-speed MRI reconstructions at the rate of OCM signals are best appreciated in video format\footnotemark[1]. \footnotetext[1]{\url{https://github.com/fpreiswerk/OCM-LRCN}}
We used publicly available code and data from \cite{Preiswerk_2016} for the KDE approach to compare the two methods, as shown by the M-mode image in Figure \ref{fig:KDE_comparison}.
In \cite{Preiswerk_2016}, a CPU reconstruction time of \SI{45}{\milli\second} per frame for a single plane was reported using KDE, on \SI{2}{\minute} of data. Using LRCN on the CPU, with the same hardware used in \cite{Preiswerk_2016}, one reconstruction took only \SI{4}{\milli\second}, for LRCN forward pass and PCA reconstruction combined. This amounts to a 10-fold speedup compared to KDE. On the GPU (NVIDIA Titan X), an additional factor of two was gained, with a reconstruction time of \SI{2}{\milli\second} (20 faster than KDE on  CPU). Moreover, the reconstruction cost of the proposed LRCN method is constant, while KDE would become even slower with increasing size of the training set. This speedup might enable multi-plane real-time image synthesis in the future. We performed a pixel-wise sum of squared error (SSE) analysis between KDE, LRCN and ground-truth images, to link our LRCN results to the quantitative validation for KDE in \cite{Preiswerk_2016}. For the dataset presented in Fig. 3.b), the average SSE per image was slightly higher with LRCN, but comparable ($39.0 \pm 12$ for LRCN vs. $33.9 \pm 7$ for KDE), which can be explained by the loss of information resulting from working in PCA subspace of the original MR images. 
In conclusion, the intriguing possibility of compressing the imaging capabilities of an MRI machine into small OCM sensors using machine learning could lead to promising image-guided therapy applications, such as real-time motion imaging for radiotherapy and biopsy needle guidance, even outside the MR bore.

\textbf{Acknowledgment.} Support from grants NIH P41EB015898, R03EB025546, R01CA149342, and R21EB019500
is duly acknowledged. GPU hardware was generously donated by NVIDIA Corporation.

\bibliography{references} 
\bibliographystyle{splncs}

\end{document}